\documentclass{article}
\usepackage{spconfa4,amsmath,graphicx}

\usepackage{comment}

\usepackage{lipsum}

\usepackage{amsfonts}
\usepackage{algorithmic}
\usepackage{algorithm}
\usepackage{array}
\usepackage[caption=false,font=normalsize,labelfont=sf,textfont=sf]{subfig}
\usepackage{textcomp}
\usepackage{stfloats}
\usepackage{url}
\usepackage{verbatim}
\usepackage{graphicx}
\usepackage[sort]{cite}
\usepackage{color}
\usepackage{xcolor}
\usepackage{tabularray}
\usepackage{tikz}
\usetikzlibrary{positioning}
\usepackage{adjustbox}
\usepackage{enumitem}   
\usepackage{pbalance}

\usepackage{booktabs}
\usepackage{hhline}
\usepackage{multirow}
\usepackage{makecell}
\usepackage{diagbox}
\usepackage{svg}
\usepackage{siunitx}
\usepackage[hidelinks]{hyperref} 

\usepackage[nolist]{acronym} 

\begin{acronym}
    \acro{df}[DF]{directivity factor}
    \acro{drr}[DRR]{direct-to-reverberant ratio}
    \acro{ndf}[NDF]{neural directional filtering}
    \acro{pesq}[PESQ]{perceptual evaluation of speech quality}
    \acro{rir}[RIR]{room impulse response}
    \acro{rtf}[RTF]{room transfer function}
    \acro{sdr}[SDR]{signal-to-distortion ratio}
    \acro{CDR}[CDR]{coherent-to-diffuse ratio}
    \acro{stft}[STFT]{short-time Fourier transform}
    \acro{vdm}[VDM]{virtual directional microphone}
    \acro{wpe}[WPE]{weighted prediction error}
    \acro{FBF}[FBF]{fixed beamformer}
	\acro{DNN}[DNN]{deep neural network}

	\acro{sdri}[$\Delta$SDR]{improvement in \ac{SDR} over the unprocessed signal}
	\acro{DMA}[DMA]{differential microphone array}
	\acro{DNN}[DNN]{deep neural network}
	\acro{DOA}[DOA]{direction-of-arrival}
	\acrodefplural{DOA}[DOAs]{directions-of-arrival}
	
	\acro{iSTFT}[iSTFT]{inverse short-time Fourier transform}
	\acro{CDMA}[CDMA]{circular \ac{DMA}}

	\acro{LDMA}[LDMA]{linear \ac{DMA}}
        \acro{LS}{least-squares}
	\acro{LSTM}[LSTM]{long short-term memory}
        \acro{BiLSTM}[BiLSTM]{bidirectional LSTM}
        \acro{UniLSTM}[UniLSTM]{unidirectional LSTM}
        \acro{WNG}[WNG]{white noise gain}
	\acro{RIRs}[RIRs]{room impulse responses}
	\acro{RIR}[RIR]{room impulse response}
        \acro{RTF}[RTF]{room transfer function}
        \acro{RTFs}[RTFs]{room transfer functions}        
	\acro{DPIR}[DPIR]{direct-path impulse response}
        \acro{MVDR}[MVDR]{minimum variance distortionless response}
        \acro{LCMV}[LCMV]{linear-constraint minimum-variance}
        \acro{PMWF}[PMWF]{parametric multichannel wiener filter}
        \acro{GSC}[GSC]{Generalized sidelobe canceller}
        \acro{FT-JNF}[FT-JNF]{joint spatial and temporal-spectral non-linear filtering}  
        \acro{JNF}[JNF]{joint non-linear filtering }        
        \acro{SSF}[SSF]{spatially selective
deep non-linear filter } 
	\acro{SDR}[SDR]{signal-to-distortion ratio}
        \acro{noisySDR}[reference microphone]{\ac{SDR} of the unprocessed omnidirectional reference microphone}
    \acrodefplural{noisySDR}[$\textrm{SDRs}^\textrm{omni}$]{\acp{SDR} of the unprocessed omnidirectional microphone}
	\acro{SNR}[SNR]{signal-to-noise ratio}
	\acro{STFT}[STFT]{short-time Fourier transform}
         \acro{MAE}[MAE]{mean absolute error}	
	\acro{TF}[TF]{time-frequency}
	\acro{tsdr}[SA-$\varepsilon$-tSDR]{source-aggregated and regularized thresholded \ac{SDR}}
      \acro{STOI}[STOI]{short term objective intelligibility}
    \acro{PESQ}[PESQ]{perceptual evaluation of speech quality}
	
	\acro{UCA}[UCA]{uniform circular array}
        \acro{NDF}[NDF]{neural directional filtering} 
        \acro{SHONDC}[SHONDC]{steerable high-order neural directional coding}
        \acro{NDSC}[NDSC]{neural directional speech coding}
        \acro{NDC}[NDC]{neural directional coding}
        \acro{WNG}[WNG]{white noise gain}
        \acro{DF}[DF]{directivity factor}
        \acro{DI}[DI]{directivity index}
        \acro{HRTF}[HRTF]{head-related transfer function}
        \acro{ILD}[ILD]{interaural level difference}
        \acro{FiLM}[FiLM]{feature-wise linear modulation}     
        \acro{PESQ}[PESQ]{perceptual evaluation of speech quality}
        \acro{UNDF}[UNDF]{neural directional filtering with user-defined directivity patterns} 
    \acro{VDM}[VDM]{virtual directional microphone}
    \acro{DirAC}[DirAC]{directional audio coding}
    \acro{FOA}[FOA]{first-order ambisonics}
    \acro{HOA}[HOA]{high-order ambisonics}
        \acro{ATF}[ATF]{acousitc transfer function}

\end{acronym}

\title{NDF+: Joint Neural Directional Filtering and Diffuse Sound Extraction}
\name{Author(s) Name(s)\thanks{Thanks to XYZ agency for funding.}}
\address{Author Affiliation(s)}

\name{Weilong Huang, Le Nhat Tam Huynh, Oliver Thiergart, Emanu{\"e}l A. P. Habets}
\address{International Audio Laboratories Erlangen\textsuperscript{$\ast$}, Am Wolfsmantel 33, 91058 Erlangen, Germany\thanks{\textsuperscript{$\ast$}A joint institution of Fraunhofer IIS and Friedrich-Alexander-Universit{\"a}t Erlangen-N{\"u}rnberg (FAU), Germany.}}

\begin{document}
\ninept
\maketitle
\begin{abstract}
Neural directional filtering (NDF) has been introduced as a flexible approach for reconstructing a virtual directional microphone (VDM) with a desired directivity pattern for spatial sound capture. Building on this idea, we propose NDF+, which enables joint neural directional filtering and diffuse sound extraction. NDF+ reformulates VDM estimation into two coupled subtasks: dereverberated VDM reconstruction and diffuse sound extraction. This reformulation enables NDF+ to manipulate diffuse components in the final reconstructed VDM output. We evaluated NDF+ under reverberant conditions and compared it with representative conventional baselines. Results show that NDF+ consistently outperforms the baselines on both subtasks while maintaining VDM reconstruction quality comparable to that of the original single-task NDF model. These findings indicate that NDF+ introduces an additional degree of freedom for diffuse sound control in the VDM reconstruction. In a stereo recording application, NDF+ provides controllable inter-channel level differences between left and right channels by adjusting the estimated diffuse component.
\end{abstract}
\section{Introduction}
\label{sec:intro}
\Ac{FBF} with an appropriate directivity pattern enables precise spatial rendering of sound sources and preserves key spatial cues, even in multi-source scenarios. However, conventional \acp{FBF}, such as \ac{DMA} \cite{benesty2012study,benesty2015design} and superdirective beamforming \cite{bitzer2001superdirective}, are fundamentally limited by a compact array with a small size and a limited number of microphones \cite{benesty2018fixed}.

Recently, \ac{NDF} has been proposed as a data-driven alternative for reconstructing a \ac{VDM} with a desired directivity pattern \cite{ndf_iwaenc, huang2025steerable, NDF}. By using a \ac{DNN} to learn the input-output behavior of an ideal directional microphone, NDF achieves a frequency-invariant target directivity pattern on compact arrays \cite{NDF}. NDF further allows flexible configuration of directivity patterns at inference \cite{huang2025neural}, making it attractive for spatial sound capture. 

Existing \ac{NDF} formulations mainly focus on \ac{VDM} reconstruction accuracy and provide limited control over diffuse components. In spatial recording scenarios, excessive diffuse sound can weaken the listener's perception of spatial cues and impair immersive audio experiences \cite{blauert1997spatial, faller2004source}, underscoring the need for effective control of diffuse components. To control the diffuseness, we propose NDF+, a joint framework for neural directional filtering and diffuse sound extraction. NDF+ reformulates VDM estimation as two coupled subtasks: (i) dereverberated VDM reconstruction and (ii) diffuse sound extraction, enabling explicit adjustment of the diffuse output. Experiments show that NDF+ consistently outperforms baselines on two specific subtasks and matches single-task NDF in the final VDM reconstruction, while uniquely allowing control over inter-channel level differences in stereo recordings via diffuse sound adjustment.

\section{Problem Formulation}
\label{sec:format}

We consider a compact array of $Q$ omnidirectional microphones recording an acoustic scene with $N$ sound sources in a reverberant room. The array and all sources are assumed to lie in the $x$-$y$ plane. Let $X_{q,n}(f,t)$ denote the \ac{STFT} coefficient at the $q$-th microphone due to the $n$-th source, where $f$ and $t$ denote the frequency-bin and time-frame indices, respectively. Under the multiplicative transfer function approximation \cite{avargel2007multiplicative}, $X_{q,n}(f,t)$ is modeled as
    \vspace*{-0.15cm}
    \begin{equation}
    \label{eqn:source_sig}
    X_{q,n}(f,t) = H_{q,n}(f) X_n(f,t), \vspace*{-0.15cm}
    \end{equation}
where $X_n(f,t)$ denotes the \ac{STFT} coefficient of the $n$-th source signal, and $H_{q,n}(f)$ is the corresponding \ac{RTF} between the $n$-th source and the $q$-th microphone. The mixture signal at the  $q$-th microphone is given by
    \vspace*{-0.15cm}
    \begin{equation} \label{eqn:mic_sig}  \vspace*{-0.15cm}
    Y_q(f,t) = \sum_{n=1}^{N} X_{q,n}(f,t) + V_q(f,t),~q\in\{1,2,\ldots,Q\}, \end{equation}
where $V_q(f,t)$ denotes sensor noise that is spatially uncorrelated across the microphones. 

The \ac{NDF} task employs a \ac{DNN} model to reconstruct a \ac{VDM} signal that captures the acoustic scene according to a desired directivity pattern \cite{ndf_iwaenc, NDF}. The \ac{VDM} is positioned at the reference microphone in the array (with $q=1$ as the reference). The directivity pattern, $\Lambda(\theta, \phi)$, defines the directional sensitivity of a beamformer or directional microphone, describing how spatial responses vary for sounds arriving from different directions \cite{elko2000superdirectional, eargle2012microphone}. Here, $\theta$ and $\phi$ represent the azimuth and polar angles of incident sound, respectively. The target \ac{VDM} signal, $Z_{\mathrm{vdm}}(f,t)$, is defined as follows:
    \vspace*{-0.15cm}
    \begin{equation}\label{eqn:vdm_sig_rvb}
        Z_{\mathrm{vdm}}(f,t)= \sum_{n=1}^{N}  
 H_{\mathrm{vdm}, n}(f; \Lambda) \, X_n(f,t), \vspace*{-0.15cm}
    \end{equation}
where $H_{\mathrm{vdm}, n}(f; \Lambda) =\sum_{i=1}^{\infty} \Lambda(\theta_i, \phi_i) \,  \rho^{(i)}_{\mathrm{vdm},n}[f ] $ denotes the \ac{RTF} between the $n$-th source and the \ac{VDM}. The term $\rho^{(i)}_{\mathrm{vdm},n}[f]$ is the transfer function of the $i$-th propagation path from the $n$-th source to the \ac{VDM} in a reverberant environment. Each reflection path is weighted by the directivity gain corresponding to its incident direction. The angles $\theta_i$ and $\phi_i$ define the incident direction for the $i$-th propagation path. The $\ac{VDM}$ signal $Z_{\mathrm{vdm}}(f,t)$ can be decomposed as follows:
    \vspace*{-0.15cm}
    \begin{equation}\label{eqn:VDMdecompose}
        Z_{\mathrm{vdm}}(f,t)= Z_{\mathrm{coh}}(f,t)+ \beta \ Z_{\mathrm{diff}}(f,t),
    \end{equation}
where $\beta = 10^{-\frac{\textrm{DI}}{20}}$ is determined by the \ac{DI} of the $\ac{VDM}$, $Z_{\mathrm{coh}}(f,t)$ denotes the spatially coherent component of the $\ac{VDM}$, and $Z_{\mathrm{diff}}(f,t)$ represents the diffuse sound captured by an omnidirectional microphone at the \ac{VDM} position. The coherent component $Z_{\mathrm{coh}}(f,t)$ is further defined as
    \vspace*{-0.15cm}
    \begin{equation}\label{eqn:VDMcoh}
       Z_{\mathrm{coh}}(f,t) = \sum_{n=1}^{N}H_{\mathrm{coh}, n}(f; \Lambda) \, X_n(f,t),
    \end{equation}
where $H_{\mathrm{coh}, n}(f; \Lambda)$ accounts for the direct sound and early reflections within $H_{\mathrm{vdm}, n}(f; \Lambda)$. The diffuse component $Z_{\mathrm{diff}}(f,t)$ is defined as
    \vspace*{-0.15cm}
    \begin{equation}\label{eqn:VDMdiff}
       Z_{\mathrm{diff}}(f,t)=\sum_{n=1}^{N} H_{\mathrm{diff}, n}(f) \, X_n(f,t),
    \end{equation}
where $H_{\mathrm{diff}, n}(f)$ corresponding to the late reverberant portion of $H_{1, n}(f)$. 

Int his work, we propose a \ac{DNN}-based method that jointly estimates $Z_{\textrm{coh}}(f,t)$ and $Z_{\mathrm{diff}}(f,t)$. Estimating ${Z}_{\mathrm{coh}}$ is equivalent to reconstructing a dereverberated \ac{VDM}, while estimating ${Z}_{\mathrm{diff}}$ corresponds to diffuse sound extraction. This approach implicitly enables the reconstruction of the \ac{VDM} target signal $Z_{\mathrm{vdm}}(f,t)$ via \eqref{eqn:VDMdecompose} for the \ac{NDF} task. Therefore, the proposed estimation process provides joint \ac{NDF} and diffuse-sound extraction.
    

\section{Proposed Method}

\subsection{DNN Architecture and Training Loss}

\begin{figure}[t!] 
\centering	
\includegraphics[width=0.599\linewidth]{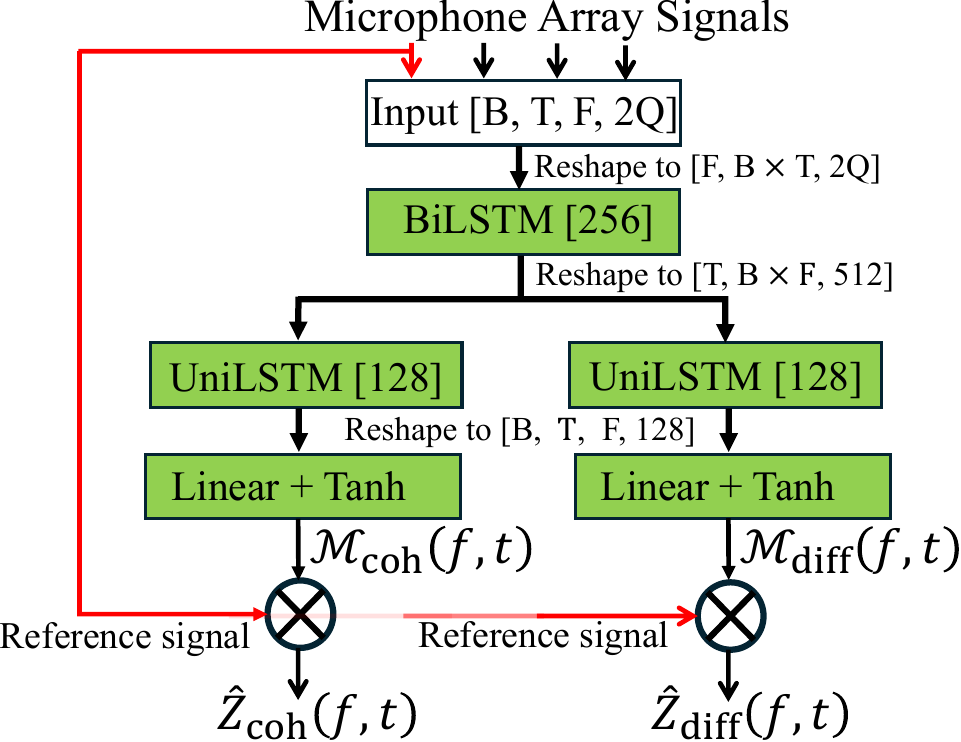}
    \vspace*{-0.15cm}
	\caption{The Dual-mask \acs{ndf} architecture. The red line represents the reference microphone signal.}
	\label{fig: dnn}
    \vspace*{-0.15cm}
\end{figure}

The FT-JNF framework \cite{FT-JNF} used for the \ac{NDF} task \cite{ndf_iwaenc, NDF} employs two distinct \ac{LSTM} networks to estimate a single complex-valued mask and applies it to a reference channel to estimate a wanted signal. To accommodate estimates for two distinct targets ($Z_{\mathrm{coh}}(f,t)$ and $Z_{\mathrm{diff}}(f,t)$), the FT-JNF is extended to a dual-mask architecture, as illustrated in Fig.~\ref{fig: dnn}. The network input remains consistent with the original FT-JNF, represented by concatenated real and imaginary components with a shape of $[B, T, F, 2Q]$, where $B$ is the batch size, $T$ is the number of time frames, and $F$ is the number of frequency bins. This input is firstly processed by a \ac{BiLSTM} network along the frequency dimension. In contrast to the single \ac{UniLSTM} used in the original FT-JNF, the proposed approach introduces two parallel \ac{UniLSTM} branches to process the \ac{BiLSTM} output along the temporal dimension. Each \ac{UniLSTM} output is then passed through a linear layer with a $\tanh$ activation to produce a complex mask. These two estimated masks ($\mathcal{M}_{\mathrm{coh}}(f,t)$ and $\mathcal{M}_{\mathrm{diff}}(f,t)$) are subsequently applied to the same reference signal, yielding two distinct estimated signals, i.e., coherent component estimate $\widehat{Z}_{\mathrm{coh}}(f,t) = \mathcal{M}_{\mathrm{coh}}(f,t) Y_{\text{1}}(f, t)$ and diffuse sound estimate $\widehat{Z}_{\mathrm{diff}}(f,t) = \mathcal{M}_{\mathrm{diff}}(f,t) Y_{\text{1}}(f, t)$. Finally, the estimated \ac{VDM} signal is obtained by: 
    \vspace*{-0.1cm}
    \begin{equation}\label{eqn:VDMdecompose_est}
     \widehat{Z}_{\mathrm{vdm}}(f,t) = \widehat{Z}_{\mathrm{coh}}(f,t) + \beta \ \widehat{Z}_{\mathrm{diff}}(f,t).
                \vspace*{-0.1cm}
    \end{equation}

We compute three losses: $\mathcal{L}_{\mathrm{coh}}$ denotes the loss between $\widehat{Z}_{\mathrm{coh}}$ and $Z_{\mathrm{coh}}$, $\mathcal{L}_{\mathrm{diff}}$ represents the loss between $\widehat{Z}_{\mathrm{diff}}$ and $Z_{\mathrm{diff}}$, and $\mathcal{L}_{\mathrm{vdm}}$ corresponds to the loss between $\widehat{Z}_{\mathrm{vdm}}$ and $Z_{\mathrm{vdm}}$. Each component is computed using a batch-aggregated normalized $\mathcal{L}_{\textrm{1}}$-loss function \cite{NDF}, defined as $\mathcal{L}_{\textrm{1}}=\frac{\sum_{b=1}^B \left \lVert \mathbf{z}^{b} - \hat{\mathbf{z}}^{b} \right \rVert_{1}}{ \sum_{b=1}^B \left \lVert  \mathbf{z}^{b} \right \rVert_{1} + \epsilon}
\vspace*{-0cm}$, where $\epsilon = 10^{-7}$. The time-domain signals $\hat{\mathbf{z}}$ and $\mathbf{z}$ correspond to the \ac{STFT} representations $\widehat{Z}$ and ${Z}$, respectively. 
The final training loss is defined as $\mathcal{L}_{\mathrm{final}} = \mathcal{L}_{\mathrm{coh}} + \mathcal{L}_{\mathrm{diff}} + \lambda_\mathrm{vdm} \; \mathcal{L}_{\mathrm{vdm}}$, where $\lambda_\mathrm{vdm} \in \{0,1\}$ is used in an ablation study.

\subsection{Training Strategy}

\begin{figure}[t]
    \centering
    
	\begin{minipage}[b]{.46\linewidth}
		\centering
		\centerline{\includegraphics[width=4cm]{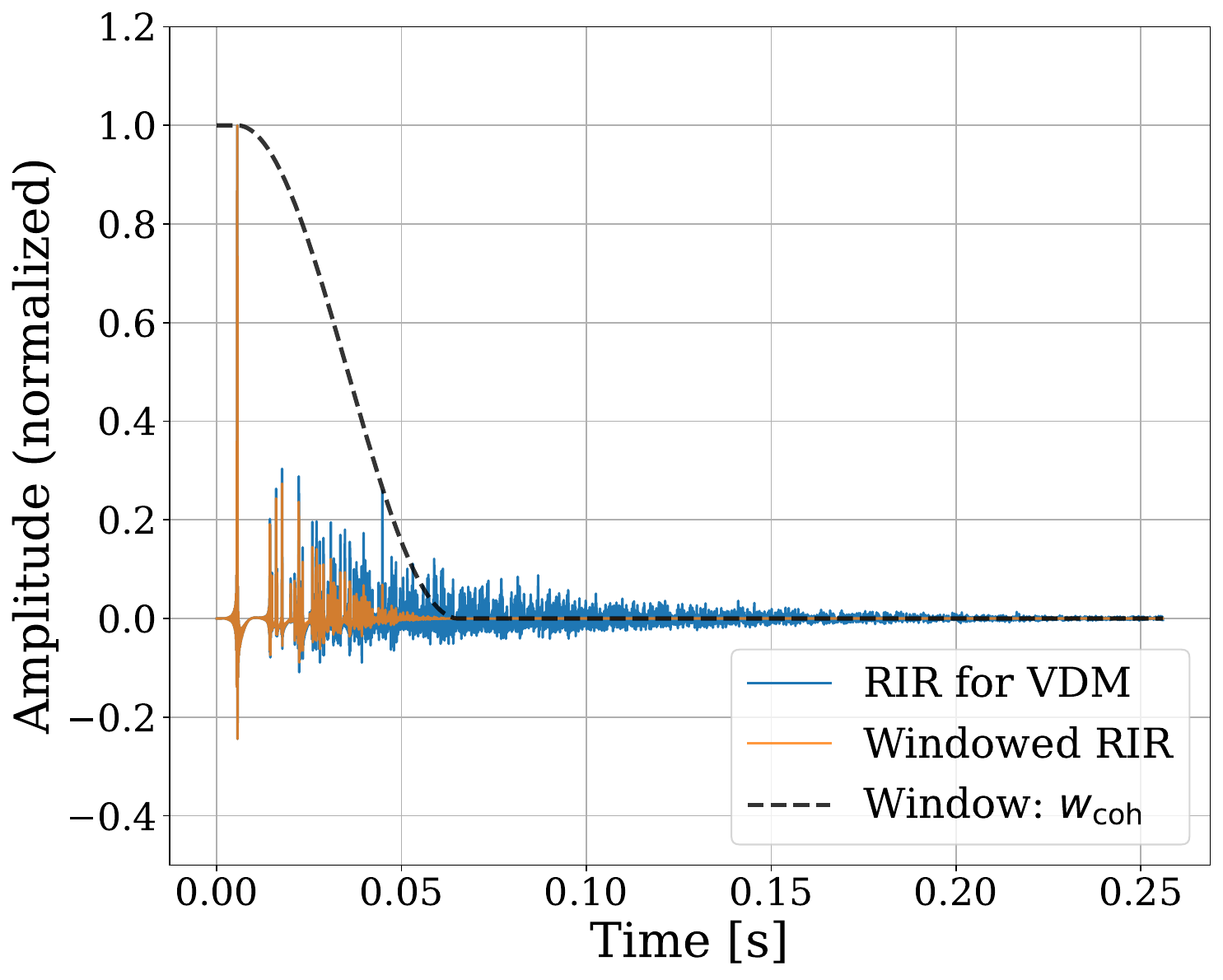}}

		\centerline{\footnotesize{(a)} }
		
	\end{minipage}
	\begin{minipage}[b]{0.46\linewidth}
		\centering
		\centerline{\includegraphics[width=4.0cm]{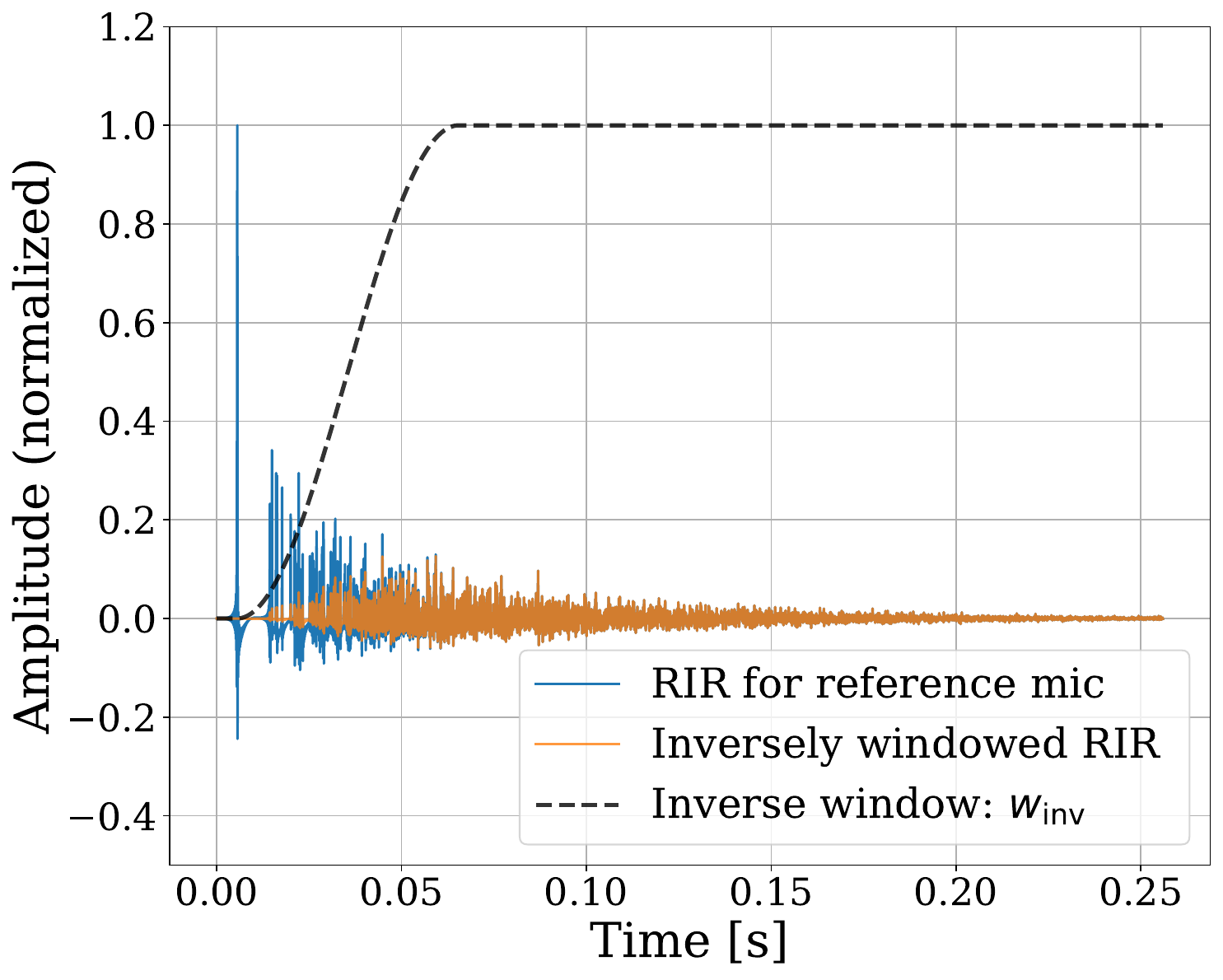}}

		\centerline{\footnotesize{(b)}  }
		
	\end{minipage}
      \vspace{-0.3cm} 		
	\caption{\small{(a) The windowing of the RIR for VDM to preserve the direct sound and early reflections; (b) the inverse windowing of the RIR for an omnidirectional microphone to extract the diffuse sound.}}
	\label{fig:window}	
     \vspace{-0.3cm}
\end{figure}

A $J^{\textrm{th}}$-order Cardioid directivity pattern \cite{NDF} is adopted as
\vspace*{-0.15cm}
\begin{equation}\label{eqn:simple_dma_pattern_definition}
\Lambda(\theta, \phi) = (0.5+ 0.5(\sin \phi \sin   \phi_\textrm{s} \cos(\theta - \theta_\textrm{s}) + \cos \phi \cos   \phi_\textrm{s} ) )^{J},
\end{equation} 
where $\theta_\textrm{s}$ and $\phi_\textrm{s}$ specify the target direction of the directivity pattern. In this study, both a $1^{\textrm{st}}$-order Cardioid with $J = 1$ and a $6^{\textrm{th}}$-order Cardioid with $J = 6$ are selected as target directivity patterns. The maximum attenuation at the null position of the directivity patterns is set to $-30$~\unit{\decibel} to ensure robust training.

 For simplicity, we assume that all microphones and sound sources lie in the $x$-$y$ plane. To learn the target directivity pattern in a reverberant room, we simulate a random source-array setup with up to three concurrent sources. The azimuth angle $\theta_{n}$ for the $n$-th speech source with respect to the array is randomly selected. Each speech source is assigned a random source-array distance. A room with random dimensions and reverberation time is defined, and the source-array setup is randomly positioned within the room. Based on the positions of the microphones and sources, the corresponding \acp{RIR} \cite{RIRGenerator} are generated, and the microphone signals are computed using \eqref{eqn:mic_sig}. 

The \ac{RIR}  for $H_{\mathrm{coh}, n}(f; \Lambda)$ is approximated by windowing the corresponding RIR of $H_{\mathrm{vdm}, n}(f; \Lambda)$, as illustrated in Fig.~\ref{fig:window}(a). The window used in Fig.~\ref{fig:window}(a) is defined as follows:
 \vspace*{-0.2cm}
\begin{equation}
w_{\text{coh}}[k] =
\begin{cases}
1, & 0 \le k < \Delta, \\
w_{\mathrm{half}}[\,k - \Delta\,], & \Delta \le k < \Delta + L, \\
0, & \Delta + L \le k < K,
\end{cases}
\end{equation}
where $\Delta$ denotes the time index of the direct-path response, $K$ is the \ac{RIR}  length, and $L$ is a constant representing the fade-out duration. The window $w_{\mathrm{half}}$, of length $L$, corresponds to the second half (fade-out portion) of a Hann window $h[k]$ of length $2L$. Then, we compute the target signal $Z_{\mathrm{coh}}(f,t)$ using \eqref{eqn:VDMcoh}. The \acp{RIR} for $H_{\mathrm{diff}, n}(f)$ is approximated by applying an inverse window to the RIR for $H_{1, n}(f)$, as shown in Fig.~\ref{fig:window}(b). The corresponding inverse window is defined as $ w_{\mathrm{inv}}[k] = 1 - w_{\mathrm{coh}}[k], \quad k = 0,1,\dots,K-1.$. Lastly, we can obtain another target signal $Z_{\mathrm{diff}}(f,t)$ using \eqref{eqn:VDMdiff}.


\section{Experimental Setup}
\label{sec:setup}

\begin{table}
        \setlength\extrarowheight{0.1pt}
        \centering
		\caption{\small{Ranges for reverberant room acoustic settings}}
		\resizebox{.38\textwidth}{!}{
			\begin{tabular}{l c rrr rrr r}
				\toprule
                   \multicolumn{1}{c}{Length} &\multicolumn{1}{c}{Width}&\multicolumn{1}{c}{Height}&\multicolumn{1}{c}{$\textrm{RT}_{60}$}&\multicolumn{1}{c}{Source-array dist.} \\
				\midrule
                     6 - 10~\unit{\metre} & 4 - 8~\unit{\metre}  & 3 - 5~\unit{\metre}  & 0.2 - 0.5~\unit{\s}  & 0.5 - 2.5~\unit{\metre}  \\

				\bottomrule
			\end{tabular}
		}
        \vspace*{-0.2cm}
		\label{tab:room_setting}
\end{table}

\noindent $\textbf{Configurations}$: A four-microphone array ($Q=4$, diameter \qty{3}{\cm}) was used, consisting of three microphones arranged in a \ac{UCA} and one positioned at the center as the reference microphone. The reference microphone signal served as the first input channel for the NDF+ model. The target direction of the directivity pattern ($\theta_s = 0$ and $\phi_s = \frac{\pi}{2}$) was assigned to a selected element of the UCA, which was designated as the second channel of input signals for the NDF+ model. The array position within the room was determined using the Monte Carlo Room Impulse Response simulation \cite{MonteCarloRIR}, ensuring a minimum distance of 1.2~\unit{\m} from all walls. The source-array distance, room size (length, width, and height), and $\textrm{RT}{60}$ were randomly sampled from the ranges specified in Table~\ref{tab:room_setting}. Speech signals for the training and validation sets were sourced from the `train-clean-360' and `dev-clean' subsets of the LibriSpeech database \cite{librispeech}, respectively. For the test sets, speech samples were drawn from the EARS dataset \cite{richter2024ears}, applying a minimum loudness threshold of $-42$dBFS \cite{loudness}. All signals were sampled at 16kHz, and $L=960$ corresponded to a 60~ms duration. The candidate incident angles for the training and validation sets were set to $\theta{n} \in {0^{\circ}, 5^{\circ}, \ldots, 355^{\circ}}$ and $\theta_{n} \in {2.5^{\circ}, 7.5^{\circ}, \ldots, 357.5^{\circ}}$. For the test set, the candidate incident angles were $\theta_{n} \in {1.25^{\circ}, 3.75^{\circ}, \ldots, 358.75^{\circ}}$. To compute $\widehat{Z}_{\mathrm{vdm}}(f,t)$, we set $\beta = 0.577$ corresponding to a $\ac{DI}=4.77$~dB for a $1^{\textrm{st}}$-order Cardioid target and $\beta = 0.277$ corresponding to a $\ac{DI} = 11.14$~dB for a $6^{\textrm{th}}$-order Cardioid target. All training stages ran for up to 150 epochs. The training set comprised 50000 samples, while the validation set included 6000 reverberant samples. Each test set contained 3240 samples with two concurrent sources. Each sample in all datasets lasted 4~s. Microphone sensor noise was added at a signal-to-noise ratio (SNR) of $30,\textrm{dB}$. The batch size was set to 10. The \ac{STFT} used a 512-point window and a 256-point hop size.

\noindent $\textbf{Performance measures}$: We used the \ac{SDR}  \cite{vincent2006performance} and \ac{PESQ} \cite{pesqc2, torcoli2025navigating} to measure distance between estimated signals and target signals. The obtained directivity patterns were estimated using the method described in \cite{NDF}.

\section{Experimental Results}

\begin{table}[t!]
\centering
\caption{Performance comparison of NDF+ and corresponding baselines for three tasks: VDM reconstruction ($\widehat{Z}_{\mathrm{vdm}}$), dereverberated \ac{VDM} reconstruction ($\widehat{Z}_{\mathrm{coh}}$), and diffuse sound extraction ($\widehat{Z}_{\mathrm{diff}}$).}
\label{tab:combined_tasks}
\sisetup{
    reset-text-series = false, 
    text-series-to-math = true, 
    mode=text,
    tight-spacing=true,
    round-mode=places,
    round-precision=2,
    table-format=2.2,
    table-number-alignment=center
}
\renewcommand{\arraystretch}{1.4}
\resizebox{0.9\columnwidth}{!}{%
\begin{tabular}{l l l c c c c c c}
    \toprule
    Task & Order & Methods & \multicolumn{2}{c}{$\mathrm{RT}_{60} = 0.2$~s} & \multicolumn{2}{c}{$\mathrm{RT}_{60} = 0.4$~s} & \multicolumn{2}{c}{$\mathrm{RT}_{60} = 0.6$~s} \\
    \cmidrule(lr){4-5} \cmidrule(lr){6-7} \cmidrule(lr){8-9}
    & & & {\acs{sdr}} & {\acs{pesq}} & {\acs{sdr}} & {\acs{pesq}} & {\acs{sdr}} & {\acs{pesq}} \\
              \midrule[1pt]
    \multirow{7}{*}{$\widehat{Z}_{\mathrm{vdm}}$} 
      & $1^{\text{st}}$ & DMA~\cite{benesty2015design} & 6.86 & 2.43 & 7.64 & 2.71 & 7.93 & 2.84 \\
      & $1^{\text{st}}$ & NDF~\cite{NDF} & \bf{22.12} & \bf{4.38} & \bf{20.37} & \bf{4.40} & \bf{19.70} & \bf{4.40} \\
      & $1^{\text{st}}$ & NDF+ (w/ $\mathcal{L}_{\mathrm{vdm}}$)& 21.42 & 4.37 & 17.98 & 4.35 & 16.44 & 4.34 \\
      & $1^{\text{st}}$ & NDF+ (w/o $\mathcal{L}_{\mathrm{vdm}}$) & 20.40 & 4.36 & 14.15 & 4.26 & 11.69 & 4.18 \\
      \cmidrule(lr){2-9}
      & $6^{\text{th}}$ & NDF~\cite{NDF} & \bf{10.58} & \bf{3.79} & \bf{7.77} & \bf{3.65} & \bf{6.92} & \bf{3.59} \\
      & $6^{\text{th}}$ & NDF+ (w/ $\mathcal{L}_{\mathrm{vdm}}$) & 10.48 & 3.77 & 7.04 & 3.50 & 5.82 & 3.36 \\
      & $6^{\text{th}}$ & NDF+ (w/o $\mathcal{L}_{\mathrm{vdm}}$) & 10.19 & 3.75 & 6.37 & 3.39 & 4.95 & 3.21 \\
    \midrule[1pt]
    \multirow{6}{*}{$\widehat{Z}_{\mathrm{coh}}$} 
            & $1^{\text{st}}$ & AWPE~\cite{4960438} + DMA & 5.22 & 2.35  & 3.03 & 2.12  & 0.98 & 1.89  \\ 
      & $1^{\text{st}}$ & DRSwWPE~\cite{huang2024practical} + DMA & 5.71 & 2.37 & 4.45 & 2.24 & 3.00 & 2.04 \\
      & $1^{\text{st}}$ & NDF+ (w/ $\mathcal{L}_{\mathrm{vdm}}$) & \bf{20.10} & 4.34 & 12.97 & 3.70 & 9.92 & 3.09 \\
      & $1^{\text{st}}$ & NDF+ (w/o $\mathcal{L}_{\mathrm{vdm}}$) & 20.03 & \bf{4.35} & \bf{13.79} & \bf{3.98} & \bf{11.09} & \bf{3.50} \\
      \cmidrule(lr){2-9}
      & $6^{\text{th}}$ & NDF+ (w/ $\mathcal{L}_{\mathrm{vdm}}$) & \bf{11.19} & 3.86 & 7.46 & 3.21 & 5.96 & 2.72 \\
      & $6^{\text{th}}$ & NDF+ (w/o $\mathcal{L}_{\mathrm{vdm}}$) & 11.08 & \bf{3.88} & \bf{7.67} & \bf{3.43} & \bf{6.44} & \bf{3.02} \\    
    \midrule[1pt]
    \multirow{5}{*}{$\widehat{Z}_{\mathrm{diff}}$} 
      & -- & Diffuse BF~\cite{diffuse_beamformer} & -13.97 & 1.85 & -2.49 & 1.99 & 0.45 & 2.09\\
      & $1^{\text{st}}$ & NDF+ (w/ $\mathcal{L}_{\mathrm{vdm}}$) & 3.77 & 2.89 & 7.02 & 3.63 & 7.96 & 3.80 \\
      & $1^{\text{st}}$ & NDF+ (w/o $\mathcal{L}_{\mathrm{vdm}}$)  & \bf{3.99} & \bf{2.96} & \bf{7.26} & \bf{3.66} & \bf{8.22} & \bf{3.82} \\
  \cmidrule(lr){2-9}
    & $6^{\text{th}}$ & NDF+ (w/ $\mathcal{L}_{\mathrm{vdm}}$) & 3.62 & 2.84 & 6.99 & 3.62 & 8.01 & 3.78 \\
      & $6^{\text{th}}$ & NDF+ (w/o $\mathcal{L}_{\mathrm{vdm}}$) & \bf{3.78} & \bf{2.86} & \bf{7.08} & \bf{3.63} & \bf{8.06} & \bf{3.80} \\    
    \bottomrule
\end{tabular}%
} \label{tab:all}
 \vspace{-0.2cm}
\end{table}

\begin{figure}[t!]
    \centering
    	\begin{minipage}[b]{0.45 \linewidth}
		\centering
		\centerline{\includegraphics[width=4.0cm]{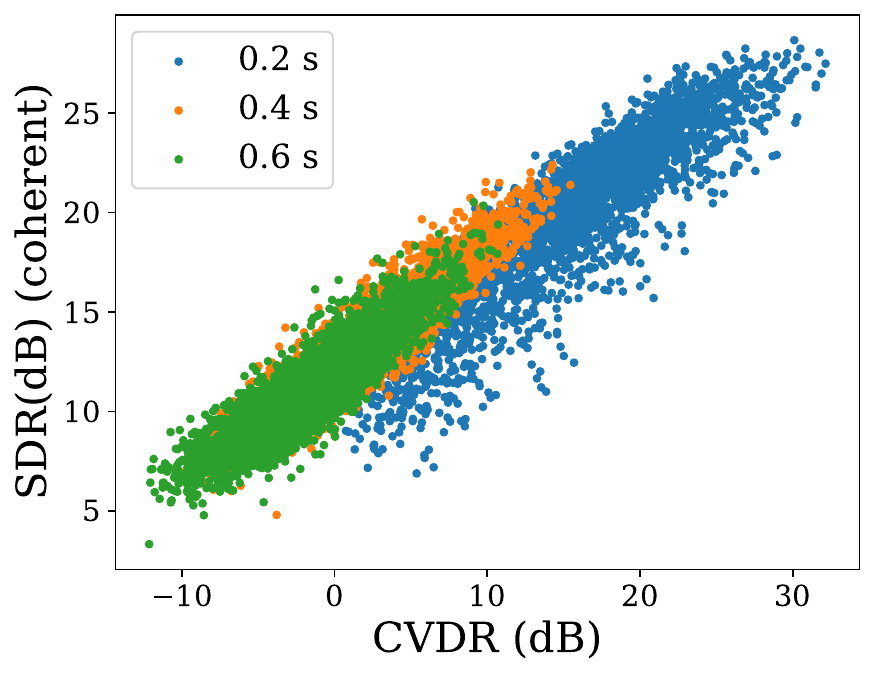}}
		(a) \small{$1^\textrm{st}$-order, $Z_{\mathrm{coh}}(f,t)$ task  }  
	\end{minipage}
	\begin{minipage}[b]{0.45\linewidth}
		\centering
		\centerline{\includegraphics[width=4.0cm]{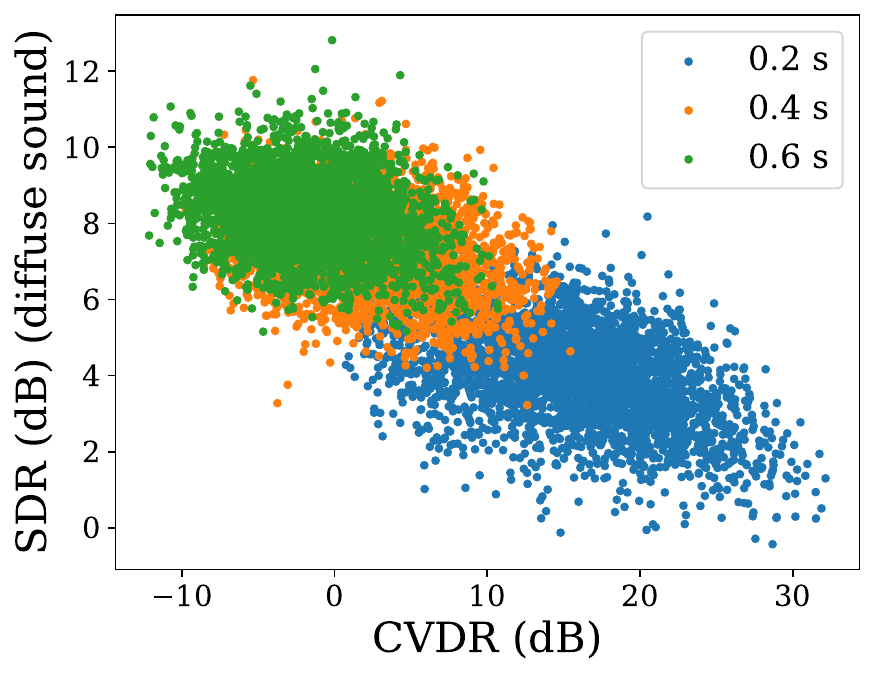}}
		(b) \small{$1^\textrm{st}$-order,  $Z_{\mathrm{diff}}(f,t)$ task }
	\end{minipage}

   	\begin{minipage}[b]{0.45\linewidth}
		\centering
		\centerline{\includegraphics[width=4.0cm]{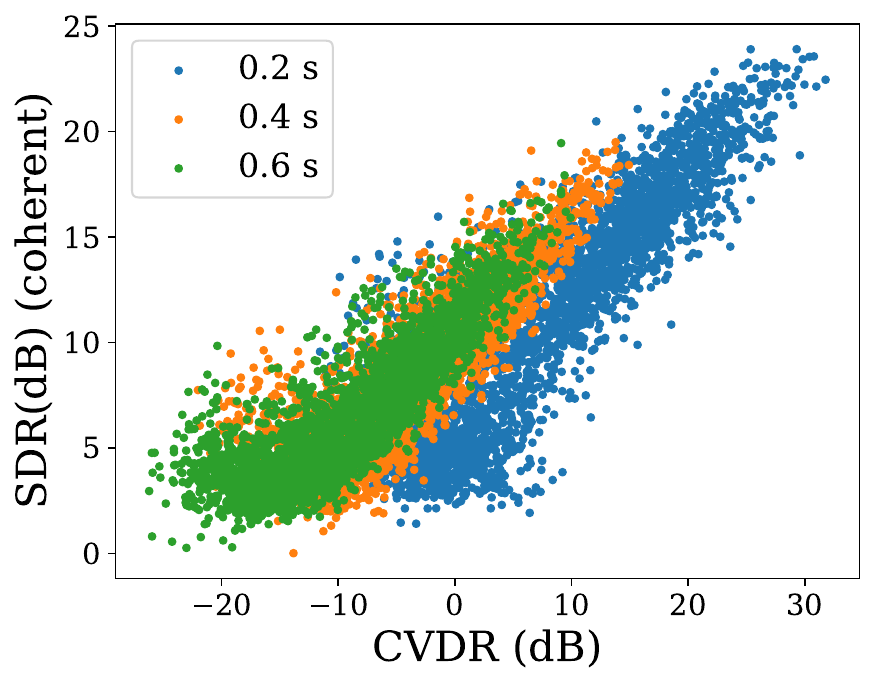}}
		(c) \small{$6^\textrm{th}$-order,   $Z_{\mathrm{coh}}(f,t)$ task  }
	\end{minipage} 
        \centering
	\begin{minipage}[b]{0.45\linewidth}
		\centering
		\centerline{\includegraphics[width=4.0cm]{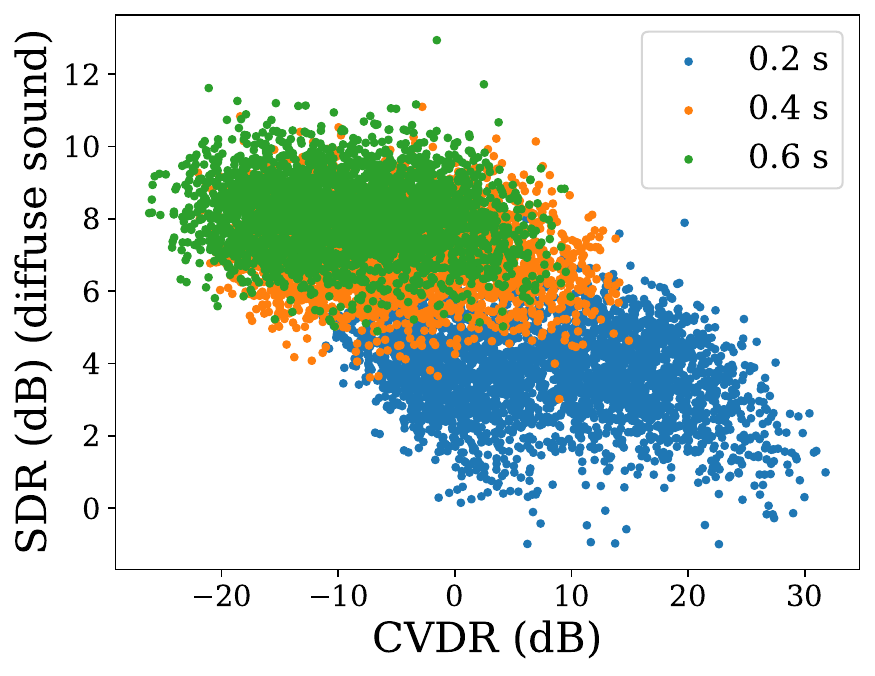}}
		(d) \small{$6^\textrm{th}$-order, $Z_{\mathrm{diff}}(f,t)$ task }
	\end{minipage}
    \vspace{-0.1cm}
    \caption{Scatter plots between the \acs{sdr} and CVDR of the NDF+ (w/o $\mathcal{L}_{\mathrm{vdm}}$) for $\mathrm{RT}_{60} \in \{0.2,0.4, 0.6\}$.}
  \label{fig:DRR_SDR_direct}
    \vspace{-0.25cm}
\end{figure}

\subsection{Performance Analysis}

\begin{figure}[t]
  \centering
    \centering
    	\begin{minipage}[b]{0.45\linewidth}
		\centering
		\centerline{\includegraphics[width=4.0cm]{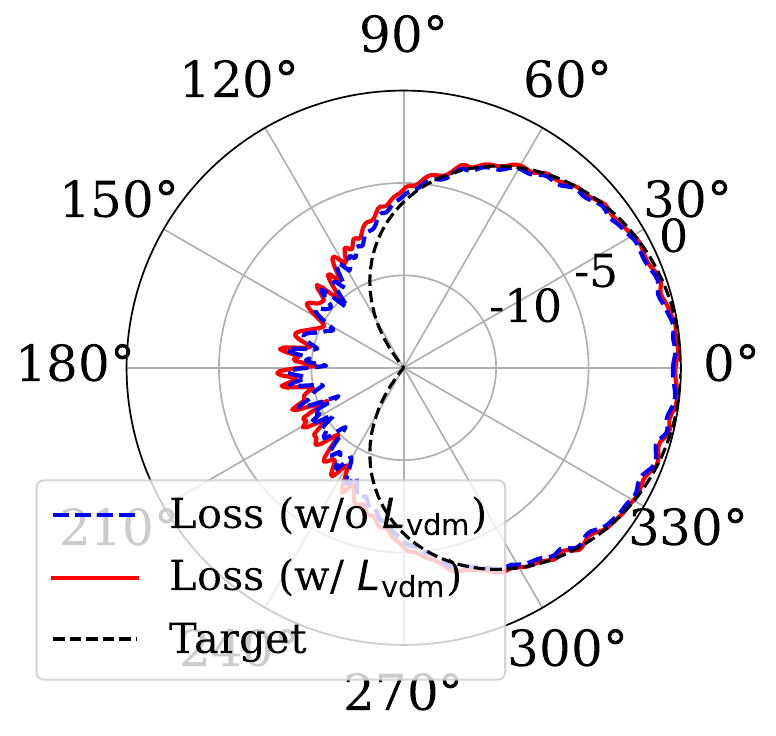}}
		(a) $1^\textrm{st}$-order, $\mathcal{M}_{\mathrm{coh}}(f,t)$  
	\end{minipage}
	\begin{minipage}[b]{0.45\linewidth}
		\centering
		\centerline{\includegraphics[width=4.0cm]{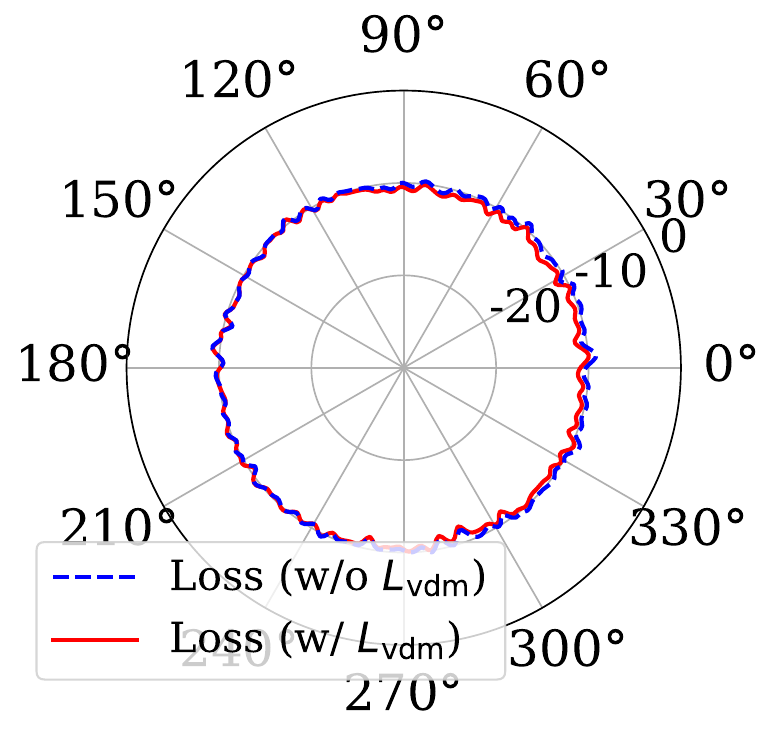}}
		(b) $1^\textrm{st}$-order, $\mathcal{M}_{\mathrm{diff}}(f,t)$ 	 
	\end{minipage}
  
   	\begin{minipage}[b]{0.45\linewidth}
		\centering
		\centerline{\includegraphics[width=4.0cm]{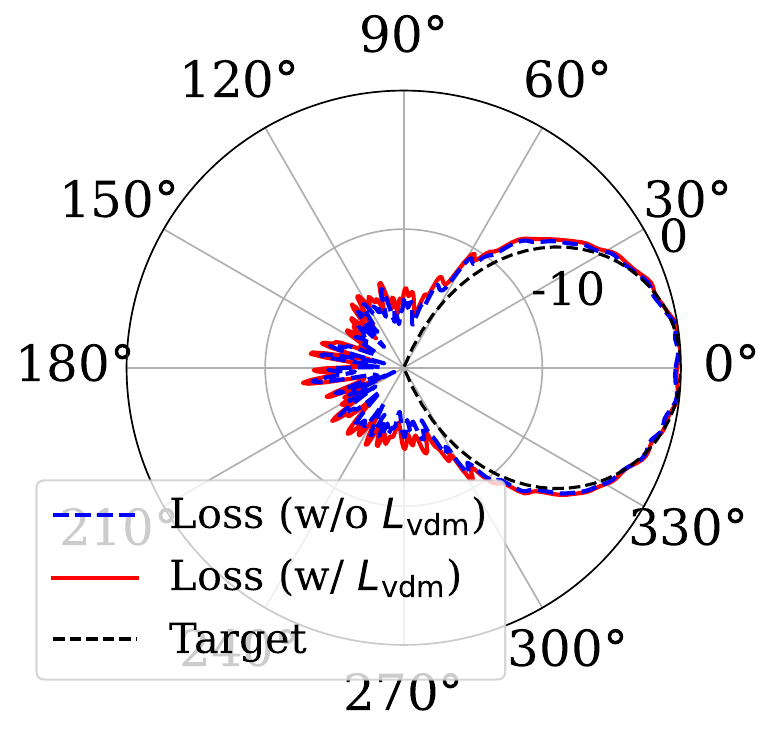}}
		(c) $6^\textrm{th}$-order, $\mathcal{M}_{\mathrm{coh}}(f,t)$   
	\end{minipage} 
        \centering
	\begin{minipage}[b]{0.45\linewidth}
		\centering
		\centerline{\includegraphics[width=4.0cm]{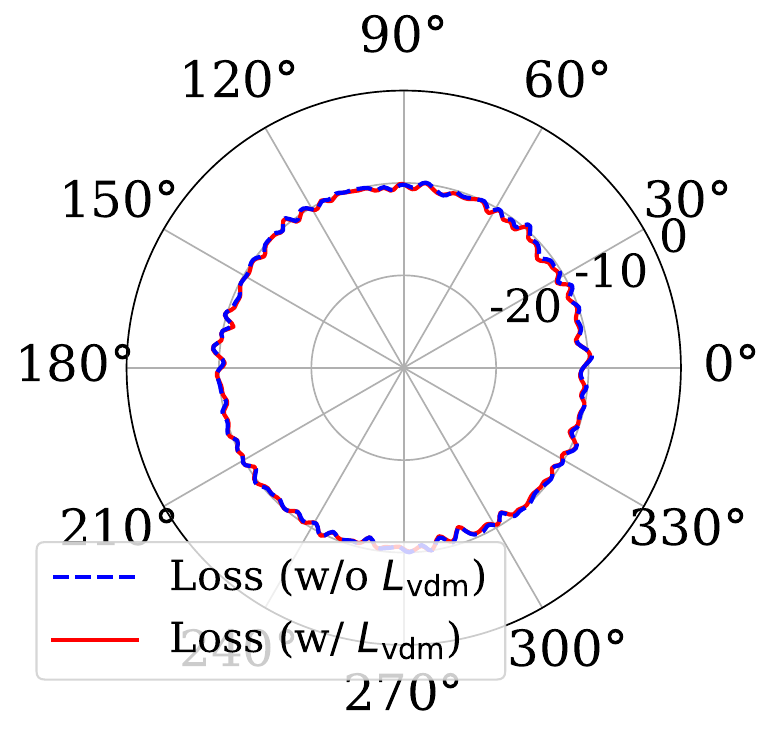}}
		(d) $6^\textrm{th}$-order, $\mathcal{M}_{\mathrm{diff}}(f,t)$
	\end{minipage}
\caption{Directivity pattern of two estimated masks $\mathcal{M}_{\mathrm{coh}}(f,t)$  and $\mathcal{M}_{\mathrm{diff}}(f,t)$ by $1^{\text{st}}$- and $6^{\text{th}}$-order NDF+ for $\mathrm{RT}_{60}=0.4$~s. }
  \label{fig:Pattern_direct_diffuse}
 \vspace{-0.35cm}
\end{figure}

The proposed NDF+ model jointly addresses two explicit subtasks: dereverberated \ac{VDM} reconstruction ($\widehat{Z}_{\mathrm{coh}}$) and diffuse sound extraction ($\widehat{Z}_{\mathrm{diff}}$). By achieving both, it implicitly realizes the \ac{VDM} reconstruction task ($\widehat{Z}_{\mathrm{vdm}}$) using \eqref{eqn:VDMdecompose_est}. Table~\ref{tab:combined_tasks} presents the results for various $\mathrm{RT}{60}$ values.

For VDM reconstruction, NDF \cite{NDF} and DMA \cite{benesty2015design} served as baselines. DMA is restricted to $1^{\text{st}}$-order and cannot operate at higher orders with the four-microphone circular array. While NDF achieves the best performance due to its task-specific optimization, NDF+, especially trained with $\mathcal{L}_{\mathrm{vdm}}$ (i.e., $\lambda_\mathrm{vdm}=1$), produces comparable results and consistently outperforms DMA. Notably, $1^{\text{st}}$-order DMA performance is degraded by noise amplification at low frequencies, attributed to low \ac{WNG}.

For dereverberated \ac{VDM} reconstruction, we built two baselines by cascading a real-time dereverberation algorithm (AWPE \cite{4960438} or DRSwWPE \cite{huang2024practical}) with DMA processing \cite{benesty2015design}. NDF+ without $\mathcal{L}_{\mathrm{vdm}}$ (i.e., $\lambda_\mathrm{vdm}=0$) achieves the best overall performance for both $1^{\text{st}}$- and $6^{\text{th}}$-order targets. For diffuse sound extraction, NDF+ outperforms the diffuse beamformer \cite{diffuse_beamformer}, whereas  NDF+ without $\mathcal{L}_{\mathrm{vdm}}$ performs best. The diffuse beamformer assumes an isotropic, homogeneous diffuse field; in less reverberant environments (e.g., $\textrm{RT}_{60} = 0.2$~s), this assumption is violated, reducing its effectiveness. Both the diffuse beamformer and NDF+ aim to suppress coherent components, which is more challenging when coherent sound dominates in low-reverberation conditions; therefore, performance improves as $\textrm{RT}_{60}$ increases. Overall, training with $\mathcal{L}_{\mathrm{vdm}}$ benefits final \ac{VDM} reconstruction for NDF+, whereas training without $\mathcal{L}_{\mathrm{vdm}}$ allows NDF+ to focus more on the two subtask estimations and yields stronger results. 

The difficulties of the two subtasks (estimation of $Z_{\mathrm{coh}}(f,t)$ and $Z_{\mathrm{diff}}(f,t)$) are inversely related: when $Z_{\mathrm{coh}}$ is much smaller than $Z_{\mathrm{diff}}$ in aggregate energy, either dereverberation or directional suppression may become more demanding, and the two may co-occur, making estimation of $Z_{\mathrm{coh}}$ harder and that of $Z_{\mathrm{diff}}$ relatively easier; the opposite holds when $Z_{\mathrm{diff}}$ is significantly smaller. To quantify this relation, we define the Coherent in VDM to Diffuse Ratio (CVDR) as $\xi = \frac{\sum_{f=1}^{F} \sum_{t=1}^{T}\left| Z_{\mathrm{coh}}(f,t) \right|^2}{\sum_{f=1}^{F} \sum_{t=1}^{T}\left| Z_{\mathrm{diff}}(f,t) \right|^2}$. We further analyzed the \ac{SDR} results in Table~\ref{tab:combined_tasks} by studying the \ac{SDR} vs. CVDR per mixture using scatter plots, as shown in Fig.~\ref{fig:DRR_SDR_direct}. On average, increasing $\textrm{RT}_{60}$ shifts the CVDR distribution toward larger values. In Figs.~\ref{fig:DRR_SDR_direct}(a) and (c), higher CVDRs lead to higher \acp{SDR} for dereverberated \ac{VDM} reconstruction ($Z_{\mathrm{coh}}(f,t)$), and NDF+ yields lower SDR with the $6^{\text{th}}$-order target than with the $1^{\text{st}}$-order target. Conversely, Figs.~\ref{fig:DRR_SDR_direct}(b) and (d) shows that lower CVDRs result in better \acp{SDR} for diffuse sound extraction ($Z_{\mathrm{diff}}(f,t)$), and $1^{\text{st}}$- and $6^{\text{th}}$-order NDF+ models are similar in scale but differ in spread. 

Figure~\ref{fig:Pattern_direct_diffuse} illustrates the directivity patterns of the two output masks for $1^{\text{st}}$- and $6^{\text{th}}$-order NDF+ models at $\mathrm{RT}_{60}=0.4$~s. The directivity pattern quantifies the masks' spatial response to direct sound from various directions \cite{NDF}. Figures~\ref{fig:Pattern_direct_diffuse}(a) and (c) show the estimated directivity for $\mathcal{M}_{\mathrm{coh}}(f,t)$, closely matching the mainlobe of the target. The model trained without $\mathcal{L}_{\mathrm{vdm}}$ achieves slightly stronger suppression around the null compared to the model trained with $\mathcal{L}_{\mathrm{vdm}}$. The $6^{\text{th}}$-order target pattern has a narrower mainlobe, resulting in a broader range of nulls, which increases approximation difficulty and lowers SDR compared to the $1^{\text{st}}$-order model. Figures~\ref{fig:Pattern_direct_diffuse}(b) and (d) display $\mathcal{M}_{\mathrm{diff}}(f,t)$, which consistently suppresses direct sound from all directions, as required for diffuse sound extraction. For diffuse sound extraction, both the $1^{\text{st}}$- and $6^{\text{th}}$-order NDF+ models yield similar directivity patterns.

\subsection{Stereo Recording Application}
\begin{figure}[t!] 
\centering	
\includegraphics[width=0.649\linewidth]{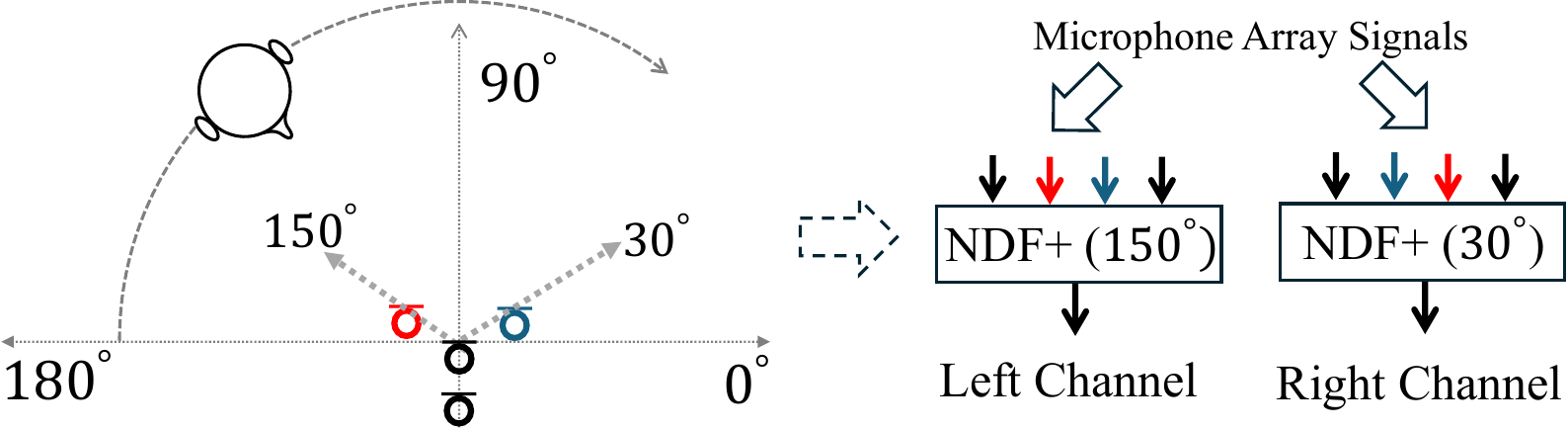}
    \vspace*{-0.1cm}
	\caption{\small{Recording setup and processing using NDF+}}
	\label{fig:setup}
    \vspace*{-0.2cm}
\end{figure}

\begin{figure}[t!] 
\centering	
\includegraphics[width=0.799\linewidth]{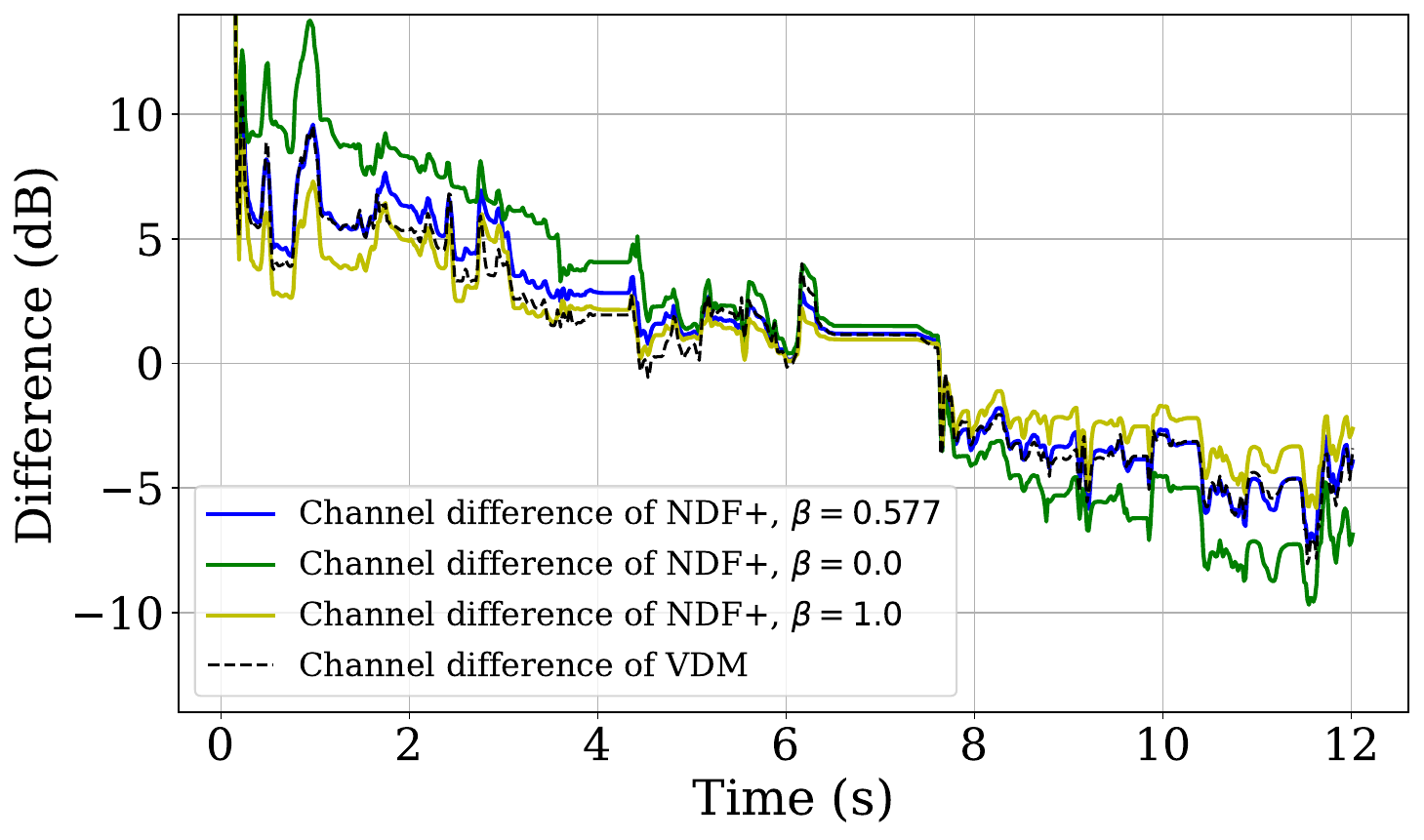}
    \vspace*{-0.4cm}
	\caption{\small{Stereo recording of NDF+ with controllable diffuse sound}}
	\label{fig: diff_channel}
    \vspace*{-0.2cm}
\end{figure}

We simulated a reverberant room ($\textrm{RT}_{60}$ = 0.5~\unit{\s}, 6~\unit{\m} $\times$ 4~\unit{\m} $\times$ 3.5~\unit{\m}). A speech source moved clockwise from $180^{\circ}$ to $0^{\circ}$ over 12 s at a fixed 1.5 m distance. Two microphones of UCA faced \qty{30}{\degree} and \qty{150}{\degree} (Fig.~\ref{fig:setup}). The scene, generated by the Dynamic Acoustic Scene Generator \cite{DASGenerator}, evaluates NDF+ stereo recording.

Stereo recording can be made with the X-Y technique using two coincident $1^{\text{st}}$-order Cardioids angled $90^{\circ}$ to $135^{\circ}$ \cite{williams2002stereophonic}. In this work, we realized the Cardiods using the NDF+ model, which was trained with a fixed target direction toward a single UCA element and $\lambda_\mathrm{vdm}=0$. By swapping the second and third input channels, the same model was steered toward either \qty{30}{\degree} or \qty{150}{\degree} (see Fig.~\ref{fig:setup}), resulting in a \qty{120}{\degree} angle. The corresponding NDF+ outputs, computed via \eqref{eqn:VDMdecompose_est} under varying $\beta$, were assigned as the left and right channels of the stereo recording. For comparison, we simulated the X-Y technique with two ideal $1^{\text{st}}$-order Cardioid \ac{VDM}s at the array center, matching the NDF+ look directions. We compared the segmental energy difference between the left and right channels for NDF+ and \ac{VDM} in Fig.~\ref{fig: diff_channel}. Results show that $\beta = 0.577$ gives NDF+ performance nearly identical to X-Y recording with ideal \acp{VDM}. Adjusting $\beta$ also allows direct control over diffuse sound energy for NDF+, thereby tuning inter-channel differences in reverberant stereo capture.

\section{Conclusions}
We introduce NDF+, a joint framework for neural directional filtering and diffuse sound extraction. NDF+ splits the VDM estimation into dereverberated VDM reconstruction and diffuse sound extraction. It consistently outperforms baselines on both tasks and matches the single-task NDF model for VDM reconstruction. Joint optimization maintains VDM reconstruction while enabling control over diffuse components. In a stereo recording in a reverberant room, this controllability allows adjustment of diffuse components and inter-channel level differences.

\pagebreak

\section{Acknowledgments}
The authors gratefully acknowledge the scientific support and HPC resources provided by the Erlangen National High Performance Computing Center (NHR@FAU) of the Friedrich-Alexander-Universität Erlangen-Nürnberg (FAU). The hardware is funded by the German Research Foundation (DFG).

\bibliographystyle{IEEEbib}
\bibliography{refs}

\end{document}